\documentstyle[preprint,prl,aps]{revtex}
\begin{document}
\draft

\title{Dependence of the BEC transition temperature 
on interaction strength: A perturbative analysis}

\author{Martin Wilkens, Fabrizio Illuminati$^{\S }$, and Meret Kr\"amer} 

\address{
Institut f\"ur Physik, Universit\"at Potsdam, \\
Am Neuen Palais 10, 14469 Potsdam, Germany \\ 
e--mail: martin.wilkens@quantum.physik.uni-potsdam.de \\
$\S $ Dipartimento di Fisica, Universit\`a di 
Salerno, and Istituto Nazionale di Fisica \\
della Materia, Unit\`a di Salerno, Via S. 
Allende, 84081 Baronissi (SA), Italy \\
e--mail: illuminati@sa.infn.it}

\date{January 28, 2000}

\maketitle

\begin{abstract}
We compute the critical temperature $T_c$ of a weakly 
interacting uniform Bose gas in the canonical ensemble,
extending the criterion of condensation provided by the counting
statistics for the uniform ideal gas.
Using ordinary perturbation theory, we find in first order 
$(T_c-T_c^0)/T_c^0 = -0.93 a\rho^{1/3}$, where $T_c^0$ is the 
transition temperature of the corresponding ideal Bose gas, 
$a$ is the scattering length, and $\rho$ is the particle 
number density. 
\end{abstract}

\vspace{0.4cm}


\section{Introduction}

It may well look like a long-solved text book excercise, but the variation of the dilute Bose gas' critical temperature with the interaction strength has not yet found a conclusive answer. To date, all authors assume a continuous behavior in the limit of weak interaction, 
$\lim_{a\rightarrow0}T_c=T_c^0$, 
where $T_c^0$ is the transition temperature of 
the non--interacting system, and $a$ is 
the s--wave scattering length.
However, the sign, the proportionality constant $c$, and 
the exponent $\eta$ in the expression for the shift in the
critical temperature at fixed density $\rho$ 
\begin{equation}
   \frac{T_c-T_c^0}{T_c^0}
   =
   \pm c
   \left[a^{3}\rho\right]^{\eta} \, ,
\end{equation}
are still subject to considerable debate. 
Early calculations by Fetter and Walecka \cite{Fetter} and Toyoda \cite{Toyoda} predict a decrease in temperature, $T_c<T_c^0$ (Yet, it should be noted that the expression derived by Fetter and Walecka yields zero for a point potential.).
However, more recent calculations indicate the opposite [3--10].
Concerning the exponent, one finds in the literature
a set of predicted rational values which range 
from $\eta=1/6$ \cite{Toyoda}, \cite{Huang}, \cite{Remark1} to $\eta=1/2$ \cite{Remark2}. 
The most recent analytical investigations converge towards 
the value $\eta=1/3$  \cite{Stoof92}, \cite{Laloe}, \cite{Baym}, \cite{Zinn},
i.~e.\ they predict a linear dependence of the temperature
shift on the scattering length. This result is
also backed by Monte Carlo simulations \cite{Gruter},
\cite{Holzmann}, and by an ingenious extrapolation of experimental data 
for the strongly interacting 
condensed He4 in vycor glass \cite{Reppy}.
Still, the result of Toyoda $\eta=1/6$ continues to 
find support \cite{Huang}. 
The proportionality constant, finally, has been predicted
to assume a variety of values, which for $\eta=1/3$ range 
from $c=0.3$ \cite{Gruter} to $c=5.1$ \cite{Reppy}. 
The Paris group most recent numerical analysis, for example, 
points at $c=2.3$ \cite{Holzmann}, a value which is 
close to the theoretical prediction of Baym and 
collaborators \cite{Zinn}, while the extrapolation of 
the experimental data on He4 in vycor glass favors $c=5.1$, 
which is closer to an early prediction $c=4.66$ of Stoof \cite{Stoof92}. 

It is frequently maintained that ordinary perturbation 
theory can not be applied as it is plagued by seemingly 
unsurmountable infrared divergencies (See \cite{Zinn}). 
We point out that this conclusion is based on the implicit 
assumption that the grand--canonical statistics, 
which is governed by a chemical potential, is a sensible 
approximation to the real system, i.~e.\ a system where -- 
as a matter of principle -- not the chemical potential, 
but rather the total number of particles is fixed, 
possibly at very large a value. 
While this assumption of thermodynamic equivalence does 
indeed hold in a system with sufficiently strong interactions, 
it must be rejected in the in the limit $a\rightarrow 0$. 
In this limit the grand--canonical statistics implies fluctuations 
of the ground state occupation, which for temperatures at and 
below $T_c$ turn out to be extravagantly large, 
$\Delta n_0\sim{\cal O}(N)$ [15--17]. It is these unphysical  
fluctuations which doom to failure any attempt to reliably 
compute the shift of the Bose gas critical temperature, 
in the non--interacting limit $a\rightarrow 0$, 
when resorting to ordinary perturbation theory in the 
grand--canonical ensemble.

As the ground state giant fluctuations are easily traced 
back to the fluctuations in the total number of particles, 
which in the grand--canonical statistics turn into an 
unacceptable $\Delta N\sim N$ for $T\leq T_c^0$, a safe 
way out is to resort to statistical ensembles where the 
total number of particles is not allowed to fluctuate. 
In the canonical and microcanonical ensembles, for example, 
the ground state fluctuations of the non--interacting system 
exhibit a scaling $\Delta n_0\sim{\cal O}(N^{2/3})$ [15] which -- 
although still anomalous -- turns out to be sufficiently 
suppressed for ordinary perturbation theory to be applicable.

Indeed, as we shall demonstrate in this letter, 
first order perturbation theory in the canonical 
ensemble yields the following shift in the critical
temperature (where $\lambda_{0}$ is the De Broglie
thermal wave length at $T = T_{c}^{0}$):
\begin{eqnarray}
   \frac{T_c-T_c^0}{T_c^0}
   & = &
   -\frac{2}{5}
   \left[\frac{8\pi}{3\zeta(3/2)}\right]
   \frac{a}{\lambda_0}
\\
   & \approx &
   -0.93\left[a^3\rho\right]^{\frac{1}{3}}\, ,
\end{eqnarray}
which -- contrary to some early expectations -- 
is neither zero nor infinite. 

\section{The Hamiltonian}

We consider a uniform system of $N$ weakly 
interacting bosons in a volume $V=L^3$, imposing
periodic boundary conditions. The Hamiltonian reads
\begin{equation}
   \hat{H} = \hat{H}_0 + \hat{H}_{\rm int}\, ,
\label{eq:H_def}
\end{equation}
where $\hat{H}_0$ is the Bose gas kinetic energy,
\begin{equation}
   \hat{H}_0 = \sum_{\bf k} \varepsilon_k \hat{n}_{\bf k}\,,
   \qquad
   \varepsilon_k = \frac{\hbar^2{\bf k}^2}{2m}\,,
\label{eq:H0_def}
\end{equation}
and $\hat{H}_{\rm int}$ describes the particle pair interaction,
\begin{equation}
   \hat{H}_{\rm int}
   =
   \frac{u}{2N}
   \sum_{{\bf pkq}}
   \hat{b}^{\dagger}_{\bf p}
   \hat{b}^{\dagger}_{\bf q}
   \hat{b}_{\bf{q}-\bf{k}}\hat{b}_{{\bf p}+{\bf k}}\,,
   \qquad
   u=\frac{4\pi\hbar^{2}a N}{mV}\,.
\label{eq:Hint_def}
\end{equation}
Here ${\bf k}=(2\pi/L){\bf n}$ is a wave vector, with ${\bf n}$ a vector of integers, $\hat{b}_{\bf k}$, $\hat{b}^{\dagger}_{\bf k}$ are 
bosonic particle annihilation and creation operators, 
$\hat{n}_{\bf k}=\hat{b}^{\dagger}_{\bf k}\hat{b}_{\bf k}$ 
is the associated number operator, $m$ denotes the particle 
mass, and $a$ denotes the s--wave scattering length. 

\section{The counting statistics}

We shall be working at fixed density $\rho=N/V$ 
(equivalently: fixed specific volume $v=\rho^{-1}$), 
but variable total number of particles $N$ (and, 
concomitantly, variable system volume $V$). 
The first issue to be faced is to provide a
definition of the transition temperature, 
which -- as we recall -- only acquires the meaning 
of a critical temperature 
in the thermodynamic limit $N\rightarrow\infty$, 
$V\rightarrow\infty$, $\rho=N/V$ constant.

We base our definition on the counting statistics 
of the zero--momentum state,
\begin{equation}
   P_n(\beta;N) = \frac{1}{Z(\beta;N)}
   {\rm Tr}
   \left\{
   \delta_{\hat{n}_{0},n}
   e^{-\beta \hat{H}}
   \delta_{\hat{N},N}
   \right\}\,,
\label{eq:Pn_def}
\end{equation}
which is the probability to find $n$ particles (out of $N$ 
total particles) in the zero--momentum state 
${\bf p}=\hbar{\bf k}=0$. Here, 
$\hat{N}=\sum_{\bf k}\hat{n}_{\bf k}$ is 
the operator for the total number of particles, 
$\delta_{a,b}$ is the Kronecker delta, 
and $Z(\beta;N)$ is the canonical partition function,
\begin{equation}
   Z(\beta;N)
   =
   {\rm Tr}
   \left\{
      e^{-\beta\hat{H}}
      \delta_{\hat{N},N}
   \right\}\,.
\label{eq:Z_def}
\end{equation}

In the non-interacting limit, the counting statistics for high temperatures is a
strictly decreasing function of $n$, i.~e.\ $P_n>P_{n+1}$ \cite{Wilkens2}. 
For sufficiently low temperatures it displays a single 
peak at $n\sim\langle n\rangle\sim O(N)$. 
Assuming that a system of weakly interacting bosons behaves correspondingly, we introduce the auxiliary function
\begin{equation}
   \tilde{D}(\beta;N)
   \equiv
   {\rm Tr}
   \left\{
      \left[\delta_{\hat{n}_0,0} - \delta_{\hat{n}_0,1}\right]
      e^{-\beta\hat{H}}
      \delta_{\hat{N},N}
   \right\}\,.
\label{eq:tildeD_def}
\end{equation}
The cross--over from the high--temperature regime, 
where $\tilde{D}>0$, to the low--temperature regime, 
where $\tilde{D}<0$, is assumed to occure for a certain value 
$\beta=\beta_*$, which is defined by the relation
\begin{equation}
   \tilde{D}(\beta_*;N) = 0\,.
\label{eq:betastar_def}
\end{equation}
For fixed density $\rho$, and fixed scattering 
length $a$, the solution of this equation depends 
on the total number of particles $N$: $\beta_*=\beta_*(N)$. 
We stipulate that, in the thermodynamic limit, 
the cross--over temperature $T_*=(k_{\rm B}\beta_*)^{-1}$ 
coincides with the critical temperature 
of Bose--Einstein condensation:
\begin{equation}
   \lim_{N\rightarrow\infty} T_*(N) = T_{\rm c}\,.
\label{eq:TstarTc_stipulation}
\end{equation}
This identification, being non--trivial for an interacting 
system, will be verified below for the non--interacting case.

\section{Perturbative analysis of $T_*$}

We determine $\beta_*$ using a series expansion 
in $\hat{H}_{\rm int}$. The Dyson series 
of $\tilde{D}=\tilde{D}(\beta;N)$ reads
\begin{equation}
   \tilde{D}=\tilde{D}_0 + \tilde{D}_1 + \tilde{D}_2 + \ldots\,,
\label{eq:tildeD_series}
\end{equation}
where $\tilde{D}_n\equiv\tilde{D}_n(\beta;N)$ is 
of $n$--th order in $\hat{H}_{\rm int}$. 
The first two terms are given by
\begin{equation}
   \tilde{D}_0(\beta;N)
   =
   {\rm Tr}\left\{
   [\delta_{\hat{n}_0,0}-\delta_{\hat{n}_0,1}]
   e^{-\beta \hat{H}_0}
   \delta_{\hat{N},N}
   \right\}\,,
\label{eq:tildeD0_def}
\end{equation}
\begin{equation}
   \tilde{D}_1(\beta;N)
   =
   -\beta
   {\rm Tr}\left\{
   \left[
   \delta_{\hat{n}_0,0}
   -
   \delta_{\hat{n}_0,1}
   \right]
   \hat{H}_{\rm int}
   e^{-\beta\hat{H}_0}
   \delta_{\hat{N},N}
   \right\}\,.
\label{eq:tildeD1_def}
\end{equation}

To solve Eq.~(\ref{eq:betastar_def}) we set
\begin{equation}
   \beta_*=\beta_*^{(0)} + \Delta\beta_*\,,
\label{eq:betastar_series}
\end{equation}
where $\beta_*^{(0)}$ denotes the cross--over 
inverse temperature of the non--interacting Bose gas, 
and $\Delta\beta_*$ is a correction which is assumed 
to be small. The defining equation for $\beta_*^{(0)}$ reads
\begin{equation}
   \tilde{D}_0(\beta_*^{(0)};N)
   =
   0\,,
\label{eq:betastar0_def}
\end{equation}
and the shift to leading order 
\begin{equation}
   \frac{\Delta\beta_*}{\beta_*^{(0)}}
   =
   \left.
   \frac{\tilde{D}_1(\beta;N)}{\beta \tilde{E}_0(\beta;N)}
   \right|_{\beta=\beta_*^{(0)}}\,,
\label{eq:deltabetastar_xpr1}
\end{equation}
where
\begin{equation}
   \tilde{E}_0(\beta;N)
   =
   -\frac{\partial}{\partial\beta}\tilde{D}_0(\beta;N)\,.
\label{eq:tildeE0_def}
\end{equation}

\subsection{Exact Relations}

Observing $\varepsilon_0=0$, which implies 
that $\hat{H}_0$ does {\em not} depend on $\hat{n}_0$, 
we may recast Eq.~(\ref{eq:tildeD0_def}) into the form
\begin{equation}
   \tilde{D}_0(\beta;N)
   =
   {\rm Tr}_{\rm ex}\left\{
   \left[\delta_{\hat{N}_{\rm ex},N}
   -\delta_{\hat{N}_{\em ex},N-1}\right]
   e^{-\beta \hat{H}_0}
   \right\}\,,
\label{eq:tildeD0_xpr1}
\end{equation}
where ${\rm Tr}_{\rm ex}$ denotes the trace over 
the occupation of excited states ${\bf k}\neq0$, and
\begin{equation}
   \hat{N}_{\rm ex} \equiv \sum_{{\bf k}\neq0} \hat{n}_{\bf k}
\label{eq:Nex_def}
\end{equation}
denotes the operator of the number of particles 
in the excited states. Furthermore, using
\begin{equation}
   \hat{H}_{\rm int}
   =
   \frac{u}{N}\hat{N}(\hat{N}-1)
   -
   \frac{u}{N}\sum_{{\bf k}}\frac{\hat{n}_{\bf 
   k}(\hat{n}_{\bf k}-1)}{2}
   +
   \hat{R}
\end{equation}
where $\hat{R}$ has no diagonal elements in the 
Fock basis, and observing that $[\delta_{\hat{n}_0,0}
-\delta_{\hat{n}_0,1}]\hat{n}_0(\hat{n}_0-1)=0$, one finds
\begin{equation}
   \tilde{D}_1(\beta;N)
   =
   \frac{u\beta}{N}
   {\rm Tr}_{\rm ex}\left\{
   \left[\delta_{\hat{N}_{\rm ex},N}
   -\delta_{\hat{N}_{\em ex},N-1}\right]
   \sum_{{\bf k}\neq0}
   \frac{\hat{n}_{\bf k}(\hat{n}_{\bf k}-1)}{2}
   e^{-\beta \hat{H}_0}
   \right\}
   -
   (N-1)u\beta \tilde{D}_0(\beta;N)\,.
\label{eq:tildeD1_xpr1}
\end{equation}
Note that due to the definition of $\beta_*^{(0)}$ 
the second terms does not contribute to the 
shift $\Delta\beta_*$. 

To proceed, we use the Laplace 
representation of the Kronecker delta
\begin{equation}
   \delta_{\hat{N}_{\rm ex},N}
   =
   \frac{1}{2\pi i}
   \int_{-i\pi}^{+i\pi}
   d\alpha
   e^{(N-\hat{N}_{\rm ex})\alpha}
\label{eq:Kronecker_xpr1}
\end{equation}
and perform the trace ${\rm Tr}_{\rm ex}$. We then face
\begin{equation}
   \tilde{D}_0(\beta;N)
   =
   \frac{1}{2\pi i}
   \int_{-i\pi}^{i\pi}
   d\alpha
   \left[1-e^{-\alpha}\right]
   e^{\tilde{F}(\alpha)}\,,
\label{eq:tildeD0_xpr2}
\end{equation}
\begin{equation}
   \tilde{D}_1(\beta;N)
   =
   \frac{u\beta}{N}
   \frac{1}{2\pi i}
   \int_{-i\pi}^{i\pi}
   d\alpha
   \left[1-e^{-\alpha}\right]
   \left(
      \sum_{{\bf k}\neq0}n^2_{\bf k}(\alpha)
   \right)
   e^{\tilde{F}(\alpha)}
   -
   (N-1)u\beta \tilde{D}_0(\beta;N)\,,
\label{eq:tildeD1_xpr2}
\end{equation}
where $n_{\bf k}(\alpha)=n_{\bf k}(\alpha;\beta,N)$, 
\begin{equation}
   n_{\bf k}(\alpha;\beta,N)
   =
   \frac{1}{e^{\alpha+\beta\varepsilon_k}-1}\,,
\label{eq:nk_def}
\end{equation}
and $\tilde{F}(\alpha)\equiv\tilde{F}(\alpha;\beta,N)$,
\begin{equation}
   \tilde{F}(\alpha;\beta,N)
   =
   N\alpha+N\frac{v}{\lambda^3}\tilde{g}_{5/2}(\alpha)\,.
\label{eq:tildeF_def}
\end{equation}
Here $\lambda\equiv\lambda(\beta)$ is the thermal 
De Broglie wave length,
\begin{equation}
   \lambda(\beta)=\sqrt{2\pi\hbar^{2}/(mk_{\rm B}T)}\,,
\label{eq:lambda_def}
\end{equation}
and $\tilde{g}_{5/2}(\alpha)\equiv
\tilde{g}_{5/2}(\alpha;\beta,N)$ 
is a discrete predecessor of a Bose integral function,
\begin{equation}
   \tilde{g}_{5/2}(\alpha;\beta,N)
   =
   - \frac{\lambda^3}{Nv}
   \sum_{{\bf k}\neq0}
   \ln\left[1-e^{-\alpha-\beta\varepsilon_k}\right]\,.
\label{eq:tildegfivehalf_def}
\end{equation}
Upon identifying $\alpha=-\beta\mu$, where $\mu$ 
denotes the chemical potential in the grand--canonical 
ensemble, we note that for fixed $\alpha$ the corresponding 
value $\tilde{F}(\alpha)$ is nothing but the ideal Bose 
gas grand--canonical free energy, and $n_{\bf k}(\alpha)$ 
is the grand--canonical mean occupation. 

\subsection{Continuum Approximation}

For sufficiently large $N$, and for the interesting 
range of thermal De Broglie wavelength such that 
$\lambda^3\rho\ll N$, we may invoke the continuum approximation, 
and replace the discrete sum over momenta by an integral, 
$\sum_{{\bf k}\neq0}\rightarrow
\frac{Nv}{(2\pi)^{3}}\int d^3k$.\footnote{In an extended 
version of this paper we shall include a systematic study of the 
corrections to the continuum approximation.} We then obtain 
\begin{eqnarray}
   \tilde{F}(\alpha;\beta,N)
   & \longrightarrow &
   F(\alpha;\beta,N)
   =
   N\alpha
   +
   N\frac{v}{\lambda^{3}}
   g_{5/2}(\alpha)\,,
\label{eq:F_def}
\\
   \sum_{{\bf k}\neq0}n^2_{\bf k}(\alpha;\beta,N)
   & \longrightarrow &
   N\frac{v}{\lambda^{3}}\left(g_{1/2}(\alpha)
   -g_{3/2}(\alpha)\right)\,,
\label{eq:sum_xpr}
\end{eqnarray}
where the $g_{\sigma}(\alpha)$ denote the Bose--Einstein 
integral functions
\begin{equation}
   g_{\sigma}(\alpha) = \frac{1}{\Gamma(\sigma)}
   \int_0^{\infty} dx
   \frac{x^{\sigma-1}}{e^{x+\alpha}-1}\,.
\label{gsigma_def}
\end{equation}
Concomitant with the above replacements we also have:
\begin{eqnarray}
   \tilde{D}_0
   & \longrightarrow &
   D_0
   =
   \int
   \frac{d\alpha}{2\pi i}
   \left[ 1 - e^{-\alpha} \right]
   e^{F(\alpha;\beta,N)}\,,
\label{eq:D0_def}
\\
   \beta \tilde{E}_0
   & \longrightarrow &
   \beta E_0
   =
   N\frac{v}{\lambda^3}
   \frac{3}{2}
   I_{5/2}(\beta;N)\,,
\label{eq:E0_def}
\\
   \tilde{D}_1
   & \longrightarrow &
   D_1
   =
   \frac{v}{\lambda^3}
   \beta u
   \left[I_{1/2}(\beta;N) - I_{3/2}(\beta;N)\right]
   -
   (N-1)\beta u D_0\,,
\label{eq:D1_def}
\end{eqnarray}
where
\begin{equation}
   I_{\sigma}(\beta;N)
   =
   \int
   \frac{d\alpha}{2\pi i}
   \left[ 1 - e^{-\alpha} \right]
   g_{\sigma}(\alpha)
   e^{F(\alpha;\beta,N)}\,.
\label{eq:Isigma_def}
\end{equation}

\subsection{Expansion in $N^{-1/3}$}

As $N$ is assumed to be large, $N\gg1$, it is now tempting
to evaluate $D_0,D_1$ in the saddle point approximation. 
Yet, due to the ${\cal O}(N^{-2/3})$ proximity of the saddle point 
to the branch point of $F$, this procedure is doomed to fail, 
and a completely different treatment must be developed.\footnote{In 
field theory, the proximity turns into a confluence which renders 
meaningless the non--interacting limit of the theory at $T=T_0$.}

Since we expect the integrals to be dominated by the 
small values of $\alpha$, we resort to the Robinson
representation \cite{Robinson}:
\begin{equation}
   g_{\sigma}(\alpha)
   =
   \Gamma(1-\sigma)\alpha^{\sigma-1}
   +
   \sum_{n=0}^{\infty}
   \frac{(-)^{n}}{n!}
   \zeta(\sigma-n)
   \alpha^{n}\,.
\label{eq:Robinson}
\end{equation}
Save for the branch cut of $\alpha^{\sigma-1}$, 
which runs along the negative real axis, the Robinson
expansion converges absolutely for $|\alpha|\leq 2\pi$. 
Note that the radius of convergence covers the domain 
of integration in the above $\alpha$--integrals.

Exploiting the Robinson representation, the exponent reads
\begin{equation}
   F
   = \ln C - Y\alpha + \frac{2}{3}X\alpha^{3/2} 
     + X{\cal O}(\alpha^2)\,,
\label{eq:F_xpr1}
\end{equation}
where $\ln C = [\zeta(5/2)/(\lambda^3\rho)]N$ is a constant, 
${\cal O}(\alpha^{2})$ denotes some analytic function, 
which may be extracted from Eq.~(\ref{eq:Robinson}), and
\begin{equation}
   X
   = \frac{2\sqrt{\pi}}{\zeta(3/2)}
     \frac{\lambda_0^3}{\lambda^3}N\,,\qquad
   Y
   = \left[\frac{\lambda_0^3}{\lambda^3} - 1 \right]N\,,
\label{eq:XY_def}
\end{equation}
with $\lambda_0^3 = \zeta(3/2)/\rho$, the thermal De Broglie 
wave length of the non--interacting gas evaluated at the
transition temperature. 

We note that for given $N$, the solution of 
Eq.~(\ref{eq:betastar_def}) implies a relation between 
$X$, which is $O(N)$, and $Y$. As we 
expect $\lambda_*\rightarrow\lambda_0$ in the 
limit $N\rightarrow\infty$, the scaling 
of $Y_*$ with $N$ is not obvious, yet
\begin{equation}
   \epsilon = \frac{Y}{X}\,
\label{eq:epsilon_def}
\end{equation}
will certainly be small. Introducing the transformation 
of the integration variable,
\begin{equation}
   \alpha\rightarrow \tau=\epsilon^{-2}\alpha
\label{eq:tau_def}
\end{equation}
the free energy reads
\begin{equation}
   F =
   \ln C
   +
   \Lambda\left(-\tau+\frac{2}{3}\tau^{3/2}\right)
   +
   \Lambda\epsilon r_{5/2}(\tau;\epsilon)\,,
\label{eq:F_xpr2}
\end{equation}
where
\begin{equation}
   \Lambda =  \frac{Y^{3}}{X^{2}}\,,
\label{eq:Lambda_def}
\end{equation}
and $r_{5/2}$ is a regular function,
\begin{equation}
   r_{5/2}(\tau;\epsilon)
   =
   \frac{\tau^{2}}{2\sqrt{\pi}}
   \sum_{\nu=0}^{\infty}
   \frac{(-\epsilon^2)^{\nu}}{(\nu+2)!}\zeta(1/2-\nu)\tau^\nu\,.
\label{eq:rfivehalf_def}
\end{equation}

Since we shall find that $\Lambda_{*}^{(0)}\sim O(1)$ at the cross--over 
temperature $T=T_{*}^{(0)}$, and concomitantly 
$\epsilon_{*}^{(0)}\sim{\cal O}(N^{-1/3})$ for large $N$, 
we may invoke a formal expansion in $\epsilon$,
\begin{equation}
   \frac{D_0}{C}
   =
   \epsilon^4 K_{1}(\Lambda)
   +
   \epsilon^{5}\frac{\zeta(1/2)}{4\sqrt{\pi}}\Lambda K_{3}(\Lambda)
   +
   {\cal O}(\epsilon^6)\,,
\label{eq:D0_epsilonserie}
\end{equation}
\begin{equation}
   \frac{I_{1/2} - I_{3/2}}{C}
   =
   \frac{\zeta(1/2)-\zeta(3/2)}{C}D_0
   +
   \epsilon^{3}\sqrt{\pi} K_{1/2}(\Lambda)
   +
   \epsilon^{4}\frac{\zeta(1/2)}{4}\Lambda K_{5/2}(\Lambda)
   +
   {\cal O}(\epsilon^{5})\,,
\label{eq:Ionehalfthreehalf_epsilonserie}
\end{equation}
\begin{equation}
   \frac{I_{5/2}}{C}
   =
   \frac{\zeta(5/2)}{C} D_0
   - \epsilon^{6}\zeta(3/2) K_{2}(\Lambda)
   - \epsilon^{7}
   \left[
   \frac{\zeta(3/2)\zeta(1/2)}{4\sqrt{\pi}}\Lambda K_4
   -
   \frac{4\sqrt{\pi}}{3} K_{5/2}
   \right]
   +
   {\cal O}(\epsilon^{8})\,,
\label{eq:Ifivehalf_epsilonserie}
\end{equation}
where we have introduced the family of functions
\begin{equation}
   K_\nu(\Lambda) = \frac{1}{2\pi i}
   \int d\tau
   \tau^{\nu}
   \exp\left\{\Lambda\left(-\tau + \frac{2}{3}\tau^{3/2}\right)\right\}\,.
\label{eq:Knu_def}
\end{equation}
The functions $K_\nu$ obey the recurrence relation
\begin{equation}
   K_{\nu + 3/2} = K_{\nu+1}-\frac{\nu+1}{\Lambda}K_{\nu}\,,
\label{eq:Knu_recurrence}
\end{equation}
which is easily proven by expressing $K$ in terms 
of $X$ and $Y$, using the inverse of 
the transformation (\ref{eq:tau_def}). 

\subsection{Results}

Upon inserting Eq.~(\ref{eq:D0_epsilonserie}) into 
Eq.~(\ref{eq:betastar0_def}), the condition which fixes 
the cross--over temperature of the non--interacting gas reads
\begin{equation}
   K_1(\Lambda_{*}^{(0)}) = 0\,.
\end{equation}
up to corrections ${\cal O}(N^{-1/3})$. 
This equation is easily solved numerically,
yielding the result $\Lambda_{*}^{(0)} = 0.334$. 
Expressed in terms of temperature we have
\begin{equation}
   T_*^{(0)}
   =
   T_c^0
   \left[
   1 
   +
   \left(
   \frac{32\pi\Lambda_*^{(0)}}{27 \zeta(3/2)^2}
   \right)^{1/3}
   \frac{1}{N^{1/3}}
   +
   {\cal O}(N^{-2/3})
   \right]
\end{equation}
where $T_c^0$ is the critical temperature of the non--interacting 
Bose gas. Note that for $N$ finite, the cross--over temperature 
is slightly higher than the transition temperature of the 
ideal Bose gas, but in the thermodynamic 
limit $N\rightarrow\infty$ they coincide. 

Collecting terms and observing that $D_0(\beta_*^{(0)};N)=0$, 
the interaction induced shift reads
\begin{equation}
   \frac{\Delta\beta_*}{\beta_*^{(0)}}
   =
   - \frac{8\pi}{3\zeta(3/2)}
   \frac{a}{\lambda_*^{(0)}}
   \left[
   \frac{K_{1/2}(\Lambda)}{\Lambda K_2(\Lambda)}
   \right]_{\Lambda=\Lambda_*^{(0)}}\,,
\label{eq:deltabetastar_xpr2}
\end{equation}
up to corrections of order ${\cal O}(N^{-1/3})$. 
The shift involves the ratio $f=K_{1/2}/(\Lambda K_{2})|_{\Lambda
=\Lambda_*}$. Exploiting the recurrence relation 
(\ref{eq:Knu_recurrence}), and observing that 
$K_{1}(\Lambda_*^{(0)})=0$, we find $f=-2/5$. 
We note that this value is exact as it does not 
depend on the numerical value of $\Lambda_*^{(0)}$. 

We are now in the position to consider the 
thermodynamic limit of Eq.~(\ref{eq:deltabetastar_xpr2}). 
Identifying $\lim_{N\rightarrow\infty}T_*=T_c$, the result reads
\begin{eqnarray}
    \frac{\Delta T_{c}}{T_{c}^{0}} \equiv 
    \frac{T_{c} - T_{c}^{0}}{T_{c}^{0}}
    & = &
    -\frac{2}{5}\frac{8\pi}{3\zeta(3/2)}\frac{a}{\lambda_0}
\\
    & = &
    -\frac{2}{5} \frac{8\pi}{3\zeta(3/2)^{4/3}}a\rho^{1/3}
    = - 0.93 a \rho^{1/3}\,.
\end{eqnarray}

We thus find a negative shift in the critical
temperature, growing linearly with the
scattering length.
This result can be compared with the other
fully analytical prediction existing in
the literature, derived by Baym, 
Blaizot and Zinn--Justin \cite{Zinn}:

\begin{equation}
\frac{\Delta T_{c}}{T_{c}^{0}} = 
\frac{8\pi}{3\zeta(3/2)^{4/3}}a\rho^{1/3} =
2.33 a \rho^{1/3} \, .
\nonumber
\end{equation}

The two results exhibit the same scaling,
but differ for the sign and for a factor $2/5$ 
in the proportionality constant.

The prediction of Baym {\it et al.} has been obtained
by evaluating the leading order in the $1/N$ expansion for
a $O(N)$ field theory model that coincides with the 
original Bose Hamiltonian for $N=2$, and by observing 
that the final result does not explicitely depend on $N$.
However, the result is strictly proven only for large
$N$, and whether it is reliable also for $N=2$ is still an open problem.

On the other hand, our approach based on 
the counting statistics and on
ordinary perturbation theory in the canonical 
ensemble indicates that in the limit
of ultraweak interaction there are 
contributions, otherwise possibly neglected in other
approaches, that tend to suppress quantum
effects. Clarification of this issue may be expected from higher order perturbation theory.

\section{Acknowledgments}
 
We would like to thank Frank Lalo\"e, Sandro
Stringari, Gordon Baym, Markus Holzmann and Maciej Lewenstein
for friendly and stimulating discussions 
in the pleasant atmosphere of the ESF 
Conference on BEC held at San Feliu de Guixol 
in the Fall 1999.

\end{document}